\documentclass{jfm}
\usepackage{graphicx}
\usepackage{epstopdf, epsfig}
\usepackage{subfigure}
\usepackage{changes}
\usepackage{amsmath}
\usepackage{siunitx}
\DeclareMathOperator{\sign}{sign}

\newcommand{\MATLABvfud}{\includegraphics[width=0.2cm]{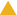}}
\newcommand{\MATLABdiamond}{\includegraphics[width=0.2cm]{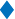}}
\newcommand{\MATLABhexagram}{\includegraphics[width=0.2cm]{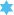}}
\newcommand{\MATLABo}{\includegraphics[width=0.2cm]{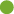}}
\newcommand{\MATLABpentagram}{\includegraphics[width=0.2cm]{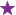}}
\newcommand{\MATLABsquare}{\includegraphics[width=0.2cm]{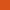}}

\shorttitle{Transition from shear-dominated to Rayleigh-Taylor turbulence}
\shortauthor{S. Brizzolara, J.-P. Mollicone, M. van Reeuwijk, A. Mazzino and M. Holzner}

\title{Transition from shear-dominated to Rayleigh-Taylor turbulence}

\author{Stefano Brizzolara\aff{1,5}
  \corresp{\email{brizzolara@ifu.baug.ethz.ch}},
  Jean-Paul Mollicone\aff{3,4},
   Maarten van Reeuwijk\aff{3},
   Andrea Mazzino\aff{2}
 \and Markus Holzner\aff{5,6}}

\affiliation{
\aff{1} Institute of Environmental Engineering, ETH Zurich, CH-8039 Zurich, Switzerland
\aff{2} DICCA, University of Genova and INFN, Genova Section, Via Montallegro 1, 16145, Genova, Italy
\aff{3} Department of Civil and Environmental Engineering, Imperial College London, London SW7 2AZ, UK
\aff{4} Division of Fluid Dynamics, Department of Mechanics and Maritime Sciences, Chalmers University of Technology, SE-41296 Gothenburg, Sweden
\aff{5} Swiss Federal Institute of Forest, Snow and Landscape Research WSL, 8903 Birmensdorf, Switzerland
\aff{6} Swiss Federal Institute of Aquatic Science and Technology Eawag, 8600 Dübendorf, Switzerland}

\begin{document}

\maketitle

\begin{abstract}
Turbulent mixing layers in nature are often characterized by the presence of a mean shear and an unstable buoyancy gradient between two streams of different velocity. Depending on the relative strength of shear versus buoyancy, either the former or the latter may dominate the turbulence and mixing between the two streams. In this paper, we  present a phenomenological theory that leads to the identification of two distinct turbulent regimes: an early regime, dominated by the mean shear, and a later regime dominated by the buoyancy. The main theoretical result consists of the identification of a cross-over time-scale that discerns between the shear- and the buoyancy-dominated turbulence. This cross-over time depends on three large-scale constants of the flow, namely the buoyancy difference, the velocity difference between the two streams, and the gravitational acceleration. We validate our theory against direct numerical simulations (DNSs) of a temporal turbulent mixing layer compounded with an unstable stratification. We observe that the cross-over time correctly predicts the transition from shear to buoyancy driven turbulence, in terms of turbulent kinetic energy production, energy spectra scaling and mixing layer thickness. 
\end{abstract}

\begin{keywords}

\end{keywords}

\section{Introduction}
Combined shear-driven and Rayleigh-Taylor (RT) turbulence occurs in mixing layers when the overlying stream is denser than the underlying one. This configuration results in a complex flow, in which the turbulent mixing process is driven both by shear and buoyancy forces. Examples of such a flow can be found in the natural environment, e.g. in the ocean or the atmosphere~\citep{turner1979buoyancy}, as well as in industrial processes, e.g. combustion chambers~\citep{nagata2000effects} and inertial confined fusion targets~\citep{atzeni2004physics}.
The two involved phenomena, i.e. RT and shear-driven turbulence, have been widely studied independently. RT turbulence phenomenology was originally introduced by \citet{chertkov2003phenomenology}, and has been recently reviewed by \citet{boffetta2017incompressible}. By shear-driven turbulence, here we refer to shear turbulent mixing layer \citep[see][Ch. 5, Sec. 5.4.2 for a complete overview]{pope2001turbulent}.

Only few works investigate the compound effect of shear and buoyancy in the mixing of two fluids. In this context, most of the past research was devoted to the analysis of the instability phase, referred to as Rayleigh-Taylor/Kelvin-Helmholtz instability (RTKHI) especially in plasma physics~\citep{finn1993nonlinear,shumlak1998mitigation,satyanarayana1984rayleigh}, but also in classical fluid flows~\citep{olson2011nonlinear}. The major result of this research is that, while a linear stability analysis predicts that adding an arbitrary shear velocity to a RT configuration will increase the perturbation growth rate, if early non-linear effects are taken into account, the shear can lead to a decrease of the growth rate, or - in extremis - to a suppression of RTKHI \citep{olson2011nonlinear, shumlak1998mitigation}.

Focusing on the fully developed turbulent phase, \citet{snider1994rayleigh} performed an experimental study in a water channel, in which an unstable thermal stratification is combined with a mean shear. The authors inferred that this problem is governed by two different types of transition: one from laminar flow to self-similar turbulence, and the other from shear-driven to buoyancy driven mixing. Focusing on the latter, the authors identified as a suitable parameter to discern between these two distinct mixing regimes the (negative) bulk Richardson number $Ri = - h g {\Delta \rho} / (\rho {\Delta U}^2)$ (where $h$ is mixing layer thickness, ${\Delta \rho}$ and $\rho$ are the density difference and the reference density respectively, ${\Delta U}$ is the shear velocity, and $g$ is the gravitational acceleration). However, their estimate for the transitional Richardson number spans a wide range, from $-5$ to $-11$. Moreover, the transition to turbulence seemed to always occur in the RT regime, and it was not possible to clearly identify the shear-dominated phase. More recent experiments performed in a gas tunnel focused on the later time of the instability phase~\citep{akula2013effect, akula2017dynamics}. The authors identified two distinct mixing regimes by the analysis of the mixing layer growth rate, that is expected to be constant for shear-dominated turbulence~\citep[][constant velocity]{pope2001turbulent}, and linear for RT turbulence~\citep[][constant acceleration]{boffetta2017incompressible}. Again, the bulk Richardson number was chosen to be the parameter to identify the transition from the shear to the RT regime. The transitional $Ri$ was shown to be within the range $(-2.5,-1.5)$ in \citep{akula2013effect} and $(-2.5,-0.8)$ in \citep{akula2017dynamics}. However, the range of investigated stratification levels is completely different from \citet{snider1994rayleigh}, which might suggest that the transitional Richardson number depends on the stratification level. Moreover, the value of $Ri$ depends sensitively on the definition of the mixing layer thickness $h$.

The complexity of the transition from shear-dominated to RT turbulence motivated investigators to devise low order Reynolds-averaged turbulence models to adequately describe it. Within this context, \citet{morgan2018two} recently proposed a two-length scale turbulence model to describe compound shear and RT mixing; within this framework, the bulk Richardson number is again employed to quantify the relative strength between shear and buoyancy forces.

Despite the effort put in this research, only few of the above mentioned works clearly capture a shear-driven, fully developed turbulent regime, since the investigated shear velocity Reynolds numbers are usually smaller or comparable than the one needed for this regime to manifest. In fact, the focus of \citet{akula2013effect, akula2017dynamics} and \citet{snider1994rayleigh} was always stated to be on the (at most) later stage of the instability phase: this is highlighted by the reference time chosen to scale the phenomena ($\sqrt{2 \rho H /(\Delta \rho g)}$) that is suitable to quantify the advancement of the RT  instability only. Another critical point regarding the above mentioned works concerns the so-called early non-linear phase of the RT instability \citep{celani2009phase, waddell2001experimental}. During the weakly non-linear phase of the RT instability, the perturbation growth rate is expected to be constant; because of this, to distinguish between the early non-linear RT and the shear-driven turbulence is extremely challenging, especially in laboratory experiments, where a large separation between the time/space at which the transition to turbulence occurs, and the time/space at which the transition from shear-dominated to RT turbulence occurs is hard to achieve. Moreover, small-scale quantities (such as the spatial velocity gradient), that could provide suitable indicators of the dominant flow regime, are difficult to access experimentally.

In this work, we investigate the fully developed turbulence arising in unstable stratified mixing layers. Our aim is to i) clarify in which conditions a transition from a shear-dominated to RT turbulence can occur, and ii) provide a phenomenological theory valid for a shear to RT transition where the shear-dominated turbulence is already developed from early times. Finally, we validate our theory against direct numerical simulations of a turbulent temporal mixing layer compounded with an unstable temperature stratification.
 
\section{Phenomenological theory \label{sec:Phenomenological theory}}
In the following, we present a phenomenological theory that describes the transition from shear-driven to RT turbulence in a temporal mixing layer, for the case when the flow is already turbulent during the shear-dominated regime. Our theory leads to the identification of a cross-over time, that discerns between the shear-dominated and the RT regimes. We assume that the fluid motion is governed by the Navier-Stokes equations in the Boussinesq approximation, i.e. density variations are neglected for inertial effects, while they are still significant in the gravitational term:
\begin{equation}
\frac{\partial \textbf{u}}{\partial t} + \textbf{u} \cdot \nabla \textbf{u} = - \frac{1}{\rho_0} \nabla p + \nu \nabla^2 \textbf{u} + \beta g \theta \textbf{e}_3,
\label{eq:NSmom}
\end{equation}
\begin{equation}
\nabla \cdot \textbf{u} = 0,
\label{eq:NScont}
\end{equation}
where $\textbf{u} = \left(u,v,w\right)$ is the fluid velocity in the stream-wise, span-wise and wall normal direction respectively, $\theta$ is the relative temperature, $\beta$ is the thermal expansion coefficient, and $g$ is the gravitational acceleration. $\nabla^2$ is the Laplace operator. The relative temperature is a scalar field that must satisfy the following advection-diffusion equation:
\begin{equation}
\frac{\partial \theta}{\partial t} + \textbf{u} \cdot \nabla \theta = \kappa \nabla^2 \theta,
\label{eq:scalar}
\end{equation}
where $\kappa$ is the thermal diffusivity.

We consider the case of a temporal turbulent mixing layer. Note that temporal mixing layers are a limit of the spatial mixing layer for which the ratio between the shear velocity $\Delta U$ (the velocity difference between the two streams) and the mean convective velocity $U_C$ (the mean velocity of the two streams) is much less than unity. In such a limit, the flow becomes statistically one-dimensional for an observer traveling in the stream-wise direction at the mean convective velocity \citep[Ch. 5, Sec. 5.4.2]{pope2001turbulent}. Such a flow was experimentally observed to hold for a velocity ratio of at least $0.60$ between the two streams \citep{bell1990development}, and was numerically reproduced using periodic boundary conditions in the stream-wise direction~\citep{rogers1994direct}. We assume that the Reynolds number is large enough so that the flow is dominated by large scale quantities only and the present turbulent state is independent on the initial perturbation and on the viscosity. Moreover, we assume that -- at early times -- the turbulence is shear-dominated, so that the balance of the NS equation at large scale gives that the first term on the LHS of (\ref{eq:NSmom}) is balanced by the nonlinear term:
\begin{equation}
\frac{u_L}{t} \sim \frac{u_L^2}{h(t)}
\end{equation}
where $u_L$ is the large-scale velocity magnitude, and $h(t)$ is the integral scale, that we identify with a measure of the mixing layer thickness. If we assume $u_L \sim \Delta U > 0$ (that is the case for shear-driven turbulence), the well known linear law for the growth of the mixing layer thickness is thus obtained:
\begin{equation}
\label{eq:KHscaling}
h(t) = S \Delta U t.
\end{equation}
The proportionality constant $S$ has been measured both experimentally and numerically, and ranges from $0.06$ to $0.11$ \citep{pope2001turbulent}. Let us now assume that the flow is gravitationally unstable, i.e. the density of the upper layer is greater than the density of the underlying one. The large scale balance becomes:
\begin{equation}
\label{eq:KHRTbalance}
\underbrace{\frac{\partial \textbf{u}}{\partial t}}_{u_L/t} + \underbrace{\textbf{u} \cdot \nabla \textbf{u}}_{u_L^2/h(t)} = ... + \underbrace{\beta g \theta \textbf{e}_3}_{\beta g \Delta \theta},
\end{equation}
where ${\Delta \theta}$ is the constant temperature difference between the two streams, and the set of dots on the RHS represents the sub-leading terms. If shear turbulence developed at early time, it must initially dominate with scaling (\ref{eq:KHscaling}). The order of magnitude of the inertial term decreases as $t^{-1}$, and thus, it follows from (\ref{eq:KHRTbalance}) that, sooner or later, the constant at the RHS must dominate. This happens when:
\begin{equation}
t \simeq t_{c} \simeq \frac{\Delta U}{\beta g {\Delta \theta}}.
\label{eq:tc}
\end{equation}
For $t \gg t_{c}$ the buoyancy term becomes dominant, thus, being always $h(t) \sim u_L \, t$, the terms on the LHS balance the buoyancy:
\begin{equation}
\frac{u_L}{t}  \sim \beta g {\Delta \theta},
\end{equation}
from which one can estimate the large scale velocity as $u_L \sim \beta g {\Delta \theta} t$. One can thus obtain the typical RT turbulence law for the mixing layer thickness as $h(t) \sim u_L t$, namely:
\begin{equation}
h(t) = \alpha \beta g {\Delta \theta} t^2,
\end{equation}
where $\alpha$ has been measured both experimentally and numerically, and ranges from $0.03$ to $0.07$ \citep{boffetta2017incompressible}.

The same analysis can be conducted from an alternative point of view. Assuming that shear initially dominates the turbulent mixing, the governing parameters are $\Delta U$ and $t$; in this phase, self-similarity leads to $h(t) \sim \Delta U t$. The ratio between buoyancy and inertial forces is described by the bulk Richardson number $Ri = \beta g {\Delta \theta} h(t) / \Delta U^2$. By imposing $Ri = 1$, one obtains $t_c = t|_{Ri = 1} = \Delta U / \beta g {\Delta \theta}$. For later times, the only relevant factors are $\beta g {\Delta \theta}$ and $t$, so that, the self-similarity of the flow implies $h(t) \sim \beta g {\Delta \theta} t^2$ and $u_L = \sqrt{\beta g {\Delta \theta} h(t)}$ which represents the free fall velocity.

When studying this flow configuration, past research focused on traditional non-dimensional quantities, namely the large scale Reynolds number $Re = u_L h(t) / \nu$, the Rayleigh number $Ra = \beta g {\Delta \theta} h(t)^3 / \left( \nu k \right)$ and the Richardson number $Ri = \beta g {\Delta \theta} h(t) / {\Delta U}^2$. All these parameters are time dependent, and are thus not suitable to describe the overall behavior of the system given a set of dimensional flow parameters (initial conditions, boundary conditions and fluid properties). By considering the functional relation between dimensional quantities, namely $h = f(\Delta U, \beta g {\Delta \theta}, k, \nu, t)$, in light of our phenomenological theory, we apply the Buckingham-$\Pi$ theorem using ${\Delta U}$ and $t_c$ as characteristics scales, and reduce the problem to the following relation between non-dimensional quantities:
\begin{equation}
\label{NDrel}
\frac{h}{h_c} = \hat{f} \left(Re_c, \, \frac{\nu}{\kappa}, \, \frac{t}{t_c} \right),
\end{equation}
where $t_c$ is the cross-over time defined in the previous section, $h_c = {\Delta U} t_c$ is the order of magnitude of the corresponding turbulent mixing layer cross-over thickness (assuming shear scaling from early times), $Re_c = {\Delta U} h_c / \nu$ is the large scale Reynolds number at the transition, and $Pr = \nu / \kappa$ is the Prandtl number. The non-dimensional mixing layer thickness, $h/h_c$, coincides with the bulk Richardson number,
\begin{equation}
\frac{h(t)}{h_c} = \frac{h(t)}{t_c \Delta U} = \frac{h(t) \beta g {\Delta \theta}}{{\Delta U}^2} = \frac{h(t) {\Delta \rho} g}{\rho {\Delta U}^2} = Ri,
\end{equation}
since $\beta {\Delta \theta} = {\Delta \rho}/\rho$. Note that, in the Boussinesq approximation $\Delta\rho / \rho = 2 A$, where $A$ is the Atwood number.

The classical mixing layer thickness Reynolds number can be expressed as $Re = Re_c \, Ri$, while the Rayleigh number is $Ra = Re_c \, Pr \, Ri$. Contrarily to the standard approach, in our perspective $Ri$ is a dependent variable. We can therefore formulate the problem in terms of the following non-dimensional variables:
\begin{equation}
t^* = \frac{t}{t_c}, \qquad \textbf{x}^* = \frac{\textbf{x}}{h_c}, \qquad \textbf{u}^* = \frac{\textbf{u}}{\Delta U}, \qquad p^* = \frac{p}{\rho_0 {\Delta U} ^2}, \qquad \theta^* = \frac{\theta}{{\Delta \theta}},
\end{equation}
which transform the momentum (\ref{eq:NSmom}), continuity (\ref{eq:NScont}) and scalar transport equations (\ref{eq:scalar}), to:
\begin{equation}
\frac{\partial \textbf{u}^*}{\partial t^*} + \textbf{u}^* \cdot \nabla \textbf{u}^* = - \nabla p^* + \frac{1}{Re_c} \nabla^2 \textbf{u}^* + \theta^* \textbf{e}_3
\label{eq:NSmomND}
\end{equation}
\begin{equation}
\nabla \cdot \textbf{u}^* = 0
\label{eq:NScontND}
\end{equation}
\begin{equation}
\frac{\partial \theta^*}{\partial t^*} + \textbf{u}^* \cdot \nabla \theta^* = \frac{1}{Re_c \, Pr} \nabla^2 \theta^*
\label{eq:scalarND}
\end{equation} 
A critical aspect that we clarify with our theory concerns the existence of the shear-dominated turbulent phase. By assuming that the shear initially dominates, we assert that our theory holds under the condition that the turbulence is already fully developed in the shear-driven regime. The reason why this is a necessary condition lies in the fact that our theory requires the global flow behavior to be governed -- from early times -- by large-scale flow parameters only, i.e. to be independent on the viscous scales. Considering equation (\ref{NDrel}), this condition requires that the cross-over Reynolds number $Re_c$ must be much larger than the shear-driven-transition-to-turbulence Reynolds number, $Re_s$. The latter is the Reynolds number at which the mixing layer reaches a fully developed turbulent state in absence of stratification. The value of $Re_s$ is documented to depend on various factors, namely: i) the velocity ratio, ii) the density ratio, iii) the initial shear layer profile, and in particular iv) the shape and nature of the instability. It can vary over a wide range ($3000 \mbox{--} 17000$)~\citep{bernal1986streamwise, koochesfahani1986mixing}. In the case of uniform density, the temporal mixing layer limit is the one for which the shear-driven-transition-to-turbulence Reynolds number is generally smaller~\citep{breidenthal1981structure}, so that, in our case, we can in principle rely on the lower limit.

\section{Direct numerical simulations}
\begin{table}
  \begin{center}
\def~{\hphantom{0}}
  \begin{tabular}{p{0.05\textwidth}p{0.05\textwidth}ccccc}
    label  & & $L_x  L_y  L_z$ & $N_x  N_y  N_z$ & $Re_c = \Delta U^3 / ( \beta g {\Delta \theta} \nu )$ & $t_c/t_0$ & $t_{end}/t_c$\\[3pt]
SL1 & \MATLABsquare & $96^3$ & $2160^3$ & $1.06 \times 10^3$ & $0.04$ & $19.63$\\
SL2 & \MATLABdiamond & $96^3$ & $2160^3$ & $4.00 \times 10^3$ & $0.15$ & $7.81$\\
SL3 & \MATLABvfud & $96^3$ & $2160^3$ & $7.99 \times 10^3$ & $0.30$ & $5.21$\\
ST4 & \MATLABpentagram  & $96^2 \times 192$ & $1080^2 \times 2160$ & $6.25 \times 10^4$  & $3.33$ & $2.30$\\
ST5 & \MATLABhexagram  & $96^2 \times 192$ & $1080^2 \times 2160$ & $6.95 \times 10^4$ & $3.58$ & $5.99$\\
ST6 & \MATLABo & $96^2 \times 192$ & $1080^2 \times 2160$ & $1.04 \times 10^5$ & $5.00$ & $1.60$\\
SNB & & $96^3$ & $2160^3$ & $\infty$ & $\infty$ & $0$\\  
  \end{tabular}
\caption{Simulations parameters: $L_i$ and $N_i$ denote the size and the number of grid points along the $i^{th}$ direction respectively; $Re_c$ is the Reynolds number corresponding to the cross-over time; $t_0$ is the time at which the flow with shear only would reach a fully developed turbulent state (i.e. $Re_{\lambda} \geq 50$ and a clear scale separation in the turbulent spectra), and $t_c$ is the cross-over time as predicted by Eq.\,(\ref{eq:tc}); $t_{end}$ is the total time of each simulation.}
  \label{tab:simParams}
  \end{center}
\end{table}
The temporal mixing layer is simulated by imposing periodic boundary conditions in the stream-wise and span-wise direction, while a free slip condition is imposed on the lower and upper walls. The stream-wise velocity profile and the scalar concentration field are initialized with a step function $u = \Delta U \sign (z) / 2 $ \citep{rogers1994direct} and $\theta  = {\Delta \theta}/2 \left(\sign (z) + 1 \right)$), respectively. Equations (\ref{eq:NSmom}), (\ref{eq:NScont}) and (\ref{eq:scalar}) are solved with a fourth-order-accurate finite volume spatial discretization scheme and a third-order Adams-Bashforth scheme for time integration \citep{craske2015energy,verstappen2003symmetry}. We fix $Pr = \nu / \kappa = 1$ in all the simulations. We opted for fixing $\Delta U = 1$ and ${\Delta \theta} = 1$, and change the control parameter $Re_c$ by acting on the kinematic viscosity $\nu$ and on the coefficient $\beta$. As shown in Sec. \ref{sec:Phenomenological theory}, $Re_c$ is indeed the only control parameter for this problem. We checked \emph{a-posteriori} that a sufficiently wide range of $t/t_c$ is explored, to be sure to observe the transition between shear-dominated and RT turbulence. In addition to the non-dimensional parameters discussed in Sec. \ref{sec:Phenomenological theory}, table \ref{tab:simParams} also shows the ratio $t_c/t_0$ where $t_0$ is the time at which the mixing layer would reach a developed turbulent state in absence of stratification, i.e. $\beta = 0$; $t_0$ is the time for which both these conditions are met: i) $Re_{\lambda} \geq 50$ and ii) a $\sim k^{-5/3}$ spectrum with at least one decade of wave number separation is visible. The first three most stratified simulations ($SL1$, $SL2$ and $SL3$) have been initialized with laminar initial conditions. In these cases, the cross-over Reynolds number is small, so that the transition to turbulence is triggered directly by RT. The time $t_0$ required for shear-dominated transition to turbulence to occur, was evaluated by performing a simulation with no buoyancy and same viscosity of $SL1$, $SL2$ and $SL3$ (i.e. $SNB$). For the three less stratified cases $ST4$, $ST5$ and $ST6$, $Re_c$ is significantly higher, meaning that the transition to turbulence is expected to occur earlier than the transition from shear to RT turbulence. In these cases we opted for switching on the buoyancy term at $t_0$, that is when the turbulence is already developed because of shear; this is equivalent to initialize the simulations with an already shear-triggered turbulent flow. This precaution, that can be realized in numerical simulations only, ensures the turbulence to be shear-dominated at $t_0$, avoiding, at the same time, the ambiguity between shear turbulent mixing and early non-linear RT instability, that characterized the previous experiments conducted with comparable shear and buoyancy forcing. The latter reasoning permits to overcome the problem of having moderately high shear Reynolds number, that characterized both experimental facilities limitations and constrains in computational power requirement of DNSs. Table \ref{tab:simParams} shows also the ratio between the total simulation time $t_{end}$ and the cross-over time $t_c$, showing that all the simulations reach the buoyancy dominated phase.

\section{\label{sec:Results  and discussion}Results and discussion}
\begin{figure}
\centering
\includegraphics[width=1\textwidth]{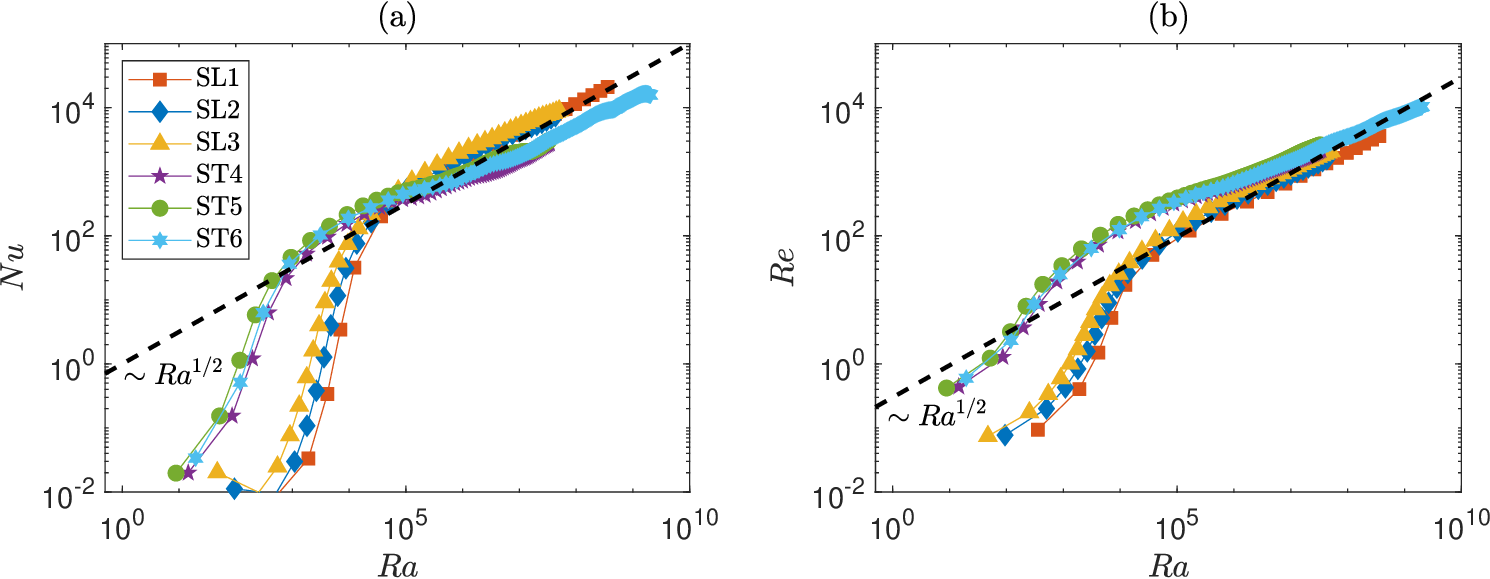}
\caption{Ultimate state scaling as a check for the turbulence to reach the  buoyancy-dominated regime; (a) shows $Nu$ vs $Ra$, while (b) shows $Re$ vs $Ra$; $Pr$ is fixed to $1$.}
\label{fig:ultimateStateScaling}
\end{figure}

To check the tendency of the flow to attain asymptotically the RT like structure, we rely on the following argument: consider the Nusselt number $Nu = \langle w' \theta' \rangle h(t) / \left( k {\Delta \theta} \right)$, namely the ratio between the total turbulent scalar transport and the diffusive transport (the brackets indicate the average within the mixing layer) and the Reynolds number $Re = u_L h(t) / \nu$. By assuming RT like scaling $u_L = \beta g \theta t$ and $h \sim \beta g {\Delta \theta} t^2$, we obtain the following relations:
\begin{equation}
Nu \sim Ra^{1/2}Pr^{1/2}, \qquad Re \sim Ra^{1/2}Pr^{-1/2},
\label{eq:ultimateScaling}
\end{equation}
where $Ra = \beta g {\Delta \theta} {h(t)}^3 / \left( \nu \kappa \right)$ is the Rayleigh number (a dimensionless measure of the density difference). This state is sometimes referred as the ultimate state of convection in the case of Rayleigh-Bénard (RB) convection \citep{kraichnan1962turbulent}; it is expected to appear at very large Rayleigh numbers, when boundary layers break down and the scalar and momentum mixing is driven by large scale contributions \citep{lohse2003ultimate}. While this regime hardly emerges in RB turbulence because of the important role played by the boundaries, in RT configurations it is well visible~\citep{boffetta2012ultimate}. Thus, these scaling laws are a good candidate to check the tendency of the flow to reach RT turbulence. From here on, we consider the temperature integral thickness $h_{\theta} = \int_{-\infty}^{+\infty} 4 \theta (1-\theta) dz$ \citep{vladimirova2009self} as a proxy of the mixing layer thickness $h(t)$. We chose this quantity instead of the momentum thickness (that is usually employed in mixing layer theory) because the latter is suitable for shear mixing layer only, while the temperature thickness is still suitable to quantify the mixing width in both limits. We recall that in our simulations, the Prandtl number $Pr = \kappa/\nu$ is fixed to $1$. Fig.~\ref{fig:ultimateStateScaling} shows clearly that all the simulated cases converge asymptotically to scaling $(Nu,Re) \sim Ra^{1/2}$, regardless of the initial stratification.

Different characteristic quantities of the flow can be analyzed to discern between shear and buoyancy driven turbulence. The most direct way to quantitatively check the validity of our theory is to analyze the turbulent kinetic energy ($tke$) balance. We chose the integral $tke$ balance (integrated over the wall normal direction), because this does not depend on the definition of the mixing layer thickness $h(t)$. The $tke$ integral balance provides:
\begin{equation}
\frac{\partial E}{\partial t} + \varepsilon = \mathcal{P}_{B} + \mathcal{P}_G 
\label{eq:tkeBalance}
\end{equation}
where $E = \int_{-\infty}^{+\infty} (\overline{\textbf{u}' \cdot \textbf{u}'})^2/2 \, \mathrm{d}z$ is the $tke$ per unit mass, $\varepsilon = \nu \int_{-\infty}^{+\infty} \overline{\left( \nabla \textbf{u}' \right)^2} \mathrm{d}z$ is the viscous dissipation, $\mathcal{P}_{B} = \beta g \int_{-\infty}^{+\infty} \overline{w' \theta'} \mathrm{d}z$ is the buoyancy production, $\mathcal{P}_{G} = - \int_{-\infty}^{+\infty} \overline{w' u'} \partial_z \overline{u} \mathrm{d}z$ is the gradient production; the over line represent the horizontal average. Initially, when the shear dominates ($t < t_c$), the $tke$ is mainly produced by $\mathcal{P}_{G}$ , i.e. the ratio between the buoyancy and the shear production is smaller than the unity. In this phase, the gradient production balances the dissipation rate ($\frac{\partial E}{\partial t} + \varepsilon \sim \mathcal{P}_G $). For later time, the buoyancy overcomes the shear, i.e. the dissipation rate is balanced by the buoyancy as in pure RT turbulence ($\frac{\partial E}{\partial t} + \varepsilon \sim \mathcal{P}_{B}$). More in detail, when $t < t_c$, we expect that $u_L \sim \Delta U$ and $h(t) \sim \Delta U t $, for both $\mathcal{P}_{B}$ and $\mathcal{P}_G$. Indeed, the buoyancy production term is expected to be passively transported by the shear velocity without any relevant influence of the buoyancy force. For later times, we expect the variation of the stream-wise mean velocity to be shear dominated. i.e. $\partial \overline{u}/\partial z  \sim \Delta U / h(t)$. The latter assertion is reasonable, since in  temporal mixing layers we have that $(\overline{v},\overline{w}) = (0,0)$, while in pure RT turbulence $\overline{u} = 0$. The stream-wise mean velocity derivative is thus expected to be mostly affected by the shear velocity scale. Concerning the wall-normal turbulent flux, we assume that $u' \sim \Delta U$ and $w'\sim \beta g \Delta \theta$, such that $\overline{u' w'}$ scales as ${\Delta U} \left( \beta g {\Delta \theta} \right) t$, i.e. the Reynolds stress tensor remains anisotropic also for longer times. We found this to agree well with our data (inset of Fig. \ref{fig:scalings} (a)). Note that this is not the case in pure RT turbulence in which the mean shear is absent. The scale of variation of the mean stream-wise velocity can be identified using any measure of the mixing layer thickness. However, it is not needed to derive the scaling of $\mathcal{P}_G$, since it drops out after integration along the wall-normal direction. Consequently, we obtain the following scalings for the shear and buoyancy $tke$ production:

\begin{figure}
\centering
\includegraphics[width = 1\textwidth]{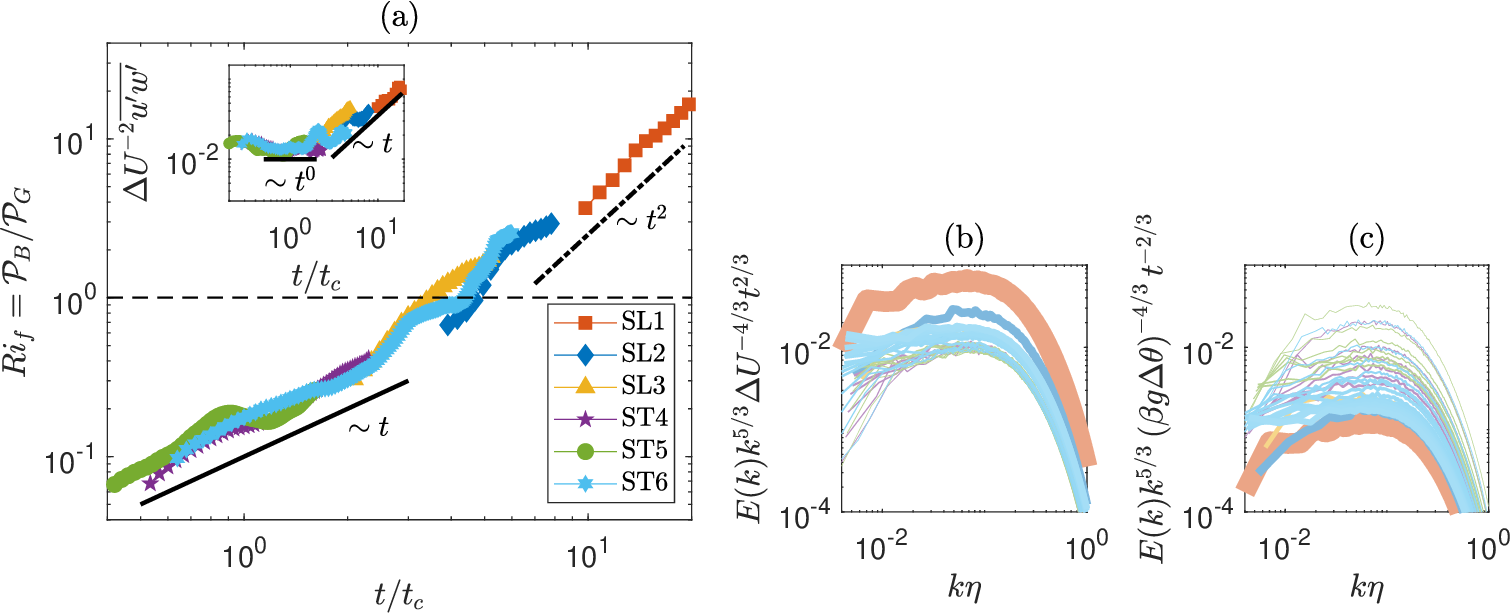}
\caption{Temporal scalings of the turbulent kinetic energy; ratio between the buoyancy and gradient $tke$ production (i.e the flux Richardson number $Ri_f$) integrated over the wall-normal direction (a); the inset shows the scaling of $\overline{u'w'}$  at the center-line; energy spectra normalized with the shear (b) and RT scalings (c) for all the simulations (see eq. \ref{eq:spectralScaling}); the line thickness is proportional to the non-dimensional time.}
\label{fig:scalings}
\end{figure}

\begin{equation}
\mathcal{P}_{G} \sim
\begin{cases}
{\Delta U}^3 &\text{if $t \ll t_c$}\\
\left( \beta g {\Delta \theta} \right) {\Delta U}^2 t &\text{if $t \gg t_c$}\\
\end{cases}
\qquad \mathcal{P}_{B} \sim
\begin{cases}
\left(\beta g {\Delta \theta}\right) {\Delta U}^2 t &\text{if $t \ll t_c$}\\
\left( \beta g {\Delta \theta} \right)^3 t^3 &\text{if $t \gg t_c$}\\
\end{cases}
\end{equation}
that implies that the flux Richardson number $Ri_f = \mathcal{P}_{B}/\mathcal{P}_{G} \sim t$ for $t \ll t_c$ and $\mathcal{P}_{B}/\mathcal{P}_{G} \sim t^2$ for $t \gg t_c$. Figure \ref{fig:scalings}\,(a) shows that our phenomenological prediction is reasonably met by the data in the limits of $t \ll t_c$ and $t \gg t_c$; as predicted, the shear-dominated phase clearly appears only in $ST4$, $ST5$ and $ST6$, for which $Re_c$ is high, and thus the transition from laminar to self-similar turbulence is shear-dominated. In the strongly stratified cases ($SL1$, $SL2$ and $SL3$), the shear-dominated turbulence is suppressed, since the buoyancy term prevails already during the transition to turbulence (low $Re_c$). A noteworthy consideration is that, even taking the simulations in which the shear-driven turbulence is inhibited, the flux-Richardson number scaling still holds, when considering only the turbulent phase.

Another argument relies on the temporal scaling of the energy spectra. By considering $K41$ inertial scaling \citep{kolmogorov1941local, kolmogorov1962refinement, obukhov1941distribution, obukhov1941spectral, obukhov1962some}, one can derive two different scaling behaviors for the second order structure function, or, equivalently, the energy spectra. In terms of energy spectra, RT and shear turbulence are expected to scale equally with the wave number $k$, but differently with time (see \citep{boffetta2017incompressible} for a complete discussion on the energy spectra temporal scaling in three dimensional RT turbulence). Simple power counting leads to the following spectral scalings:
\begin{equation}
\label{eq:spectralScaling}
E(k) \sim
\begin{cases}
\Delta U^{4/3} t^{-2/3} k^{-5/3} &\text{if $t \ll t_c$}\\
\left( \beta g {\Delta \theta} \right)^{4/3} t^{2/3} k^{-5/3} &\text{if $t \gg t_c$}\\
\end{cases}
\end{equation}
Fig. \ref{fig:scalings} (b) and (c) show that shear/RT scaling holds for early/late times respectively. Indeed, all the curves tend to stabilize on an RT like spectra for later non-dimensional times (thick line), and the shear scaling holds for early non-dimensional times (thin line), while becomes not reliable for later non-dimensional times (thicker line).

\begin{figure}
\centering
\includegraphics[width = 1\textwidth]{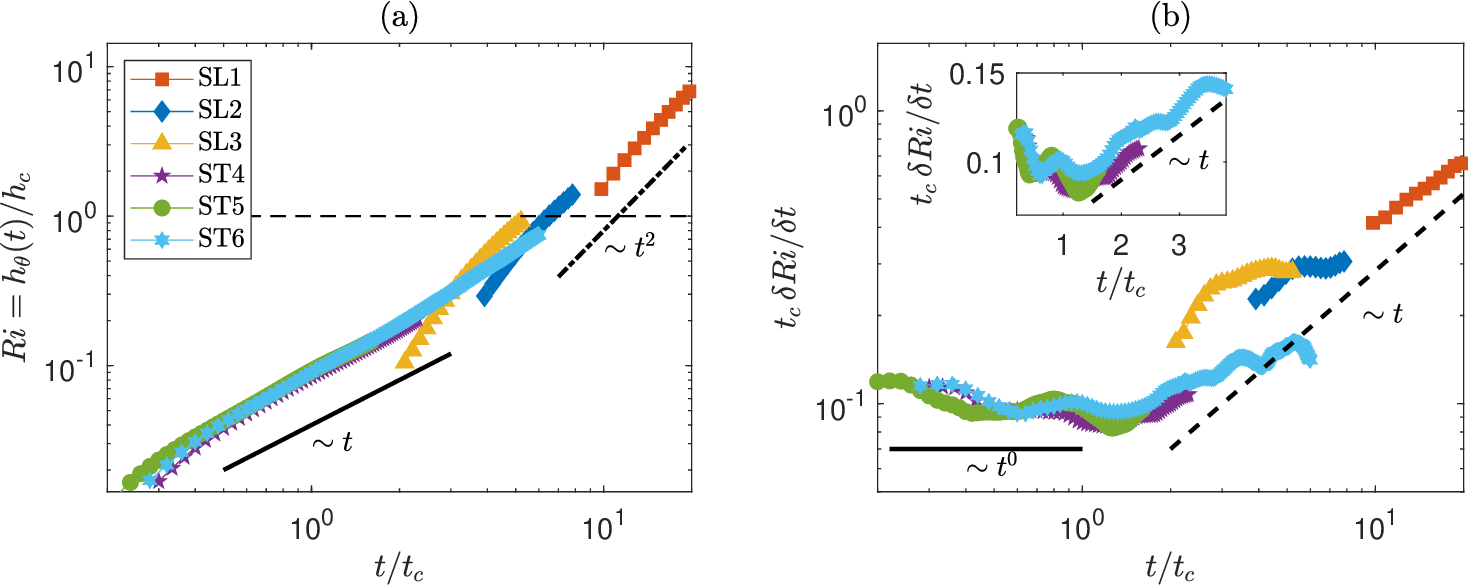}
\caption{Bulk Richardson number (a), and bulk Richardson number growth rate (b); the inset shows $ST4$, $ST5$ and $ST6$ in linear space; note that the bulk Richardson number growth rate is a non-dimensional measure of the mixing layer thickness growth rate, since $h_{\theta}$ is proportional to $Ri$.}
\label{fig:Ri_and_increments}
\end{figure}

Finally, the two regimes can also be distinguished by analyzing the bulk Richardson number $h(t)/h_c$ and the mixing layer growth rate $\delta h(t) / \delta t$. In the shear-dominated phase, a linear spreading of the mixing layer is expected, while when buoyancy prevails, the mixing region grows quadratically. In other words, the time growth rate is expected to be constant in the shear-dominated phase and linear in the RT phase. Despite the fact that this method is the one usually adopted to distinguish between the two regimes, it does often not provide a clear evidence of the dominance of the shear on the buoyancy and viceversa. The reason is that the tendency of the mixing layer to grow linearly or quadratically is not a direct measure of the predominance of one forcing on the other, but only a consequence, i.e. not directly causally correlated to the phenomena (in support of this argument, one may consider the early non-linear phase of RT). Fig. \ref{fig:Ri_and_increments} (a) shows that, as expected, $Ri$ grows linearly for $t \ll t_c$ and quadratically for $t \gg t_c$. Fig. \ref{fig:Ri_and_increments} (b) shows $\delta h(t) / \delta t$: for $ST4$, $ST5$ and $ST6$, an initial constant-like behavior that becomes linear around $t = t_c$ is appreciable, while $SL1$, $SL2$ and $SL3$ shows only the linear trend, meaning that the shear-dominated turbulent phase is suppressed. The inset of Fig.\ref{fig:Ri_and_increments} (b), shows the growth rate in linear space for $ST4$, $ST5$ and $ST6$, highlighting the transitional phase between shear-dominated and RT turbulence.

Having shown that the cross-over time $t_c$ correctly scales the bulk and the flux Richardson number, one may be interested in the exact value of the non-dimensional time at which the transition effectively occurs, termed $t/t_c|_T$. To estimate $t/t_c|_T$ we adopt the following procedure for $Ri$, $Ri_f$ and $t_c \delta {Ri}/ {\delta t}$. Firstly, we fit the phenomenological asymptotic prediction for $t \ll t_c$ and $t \gg t_c$ into the early and late time data of $Ri$, $Ri_f$ and $t_c \delta {Ri}/ {\delta t}$. This is done systematically, by increasing the distance from $t/t_c = 1$ since the constant fitting coefficients do not change any more. Then, the transitional time is estimated as the intersection between the two fitted asymptotic laws. For instance, by considering the bulk Richardson number, we expect $Ri = b_{shear} \, \left( t/t_c \right)^1$ at early times and $Ri = b_{RT} \, \left( t/t_c \right)^2$ for later times; $b_{shear}$ and $b_{RT}$  are evaluated through the above-mentioned fitting procedure, and the transitional time is obtained as $t/t_c|_T = b_{shear}/b_{RT}$. Tab. \ref{tab:transition} shows the values of $t/t_c|_T$ for each of the considered observables, and their corresponding value at each of the transitional times. We report also the value of the gradient Richardson number evaluated at the center-line ($Ri_g = -\frac{g}{\rho} \frac{\partial \overline{\rho}}{\partial z} / \left(\frac{\partial \overline{u}}{\partial z}\right)^2$ at $z = 0$) and the corresponding transitional time.

\begin{table}
  \begin{center}
\def~{\hphantom{0}}
  \begin{tabular}{lcccc}
								& $t/t_c|_T$   & $Ri(t/t_c|T)$ & $Ri_f(t/t_c|T)$ & $Ri_g(t/t_c|T, z = 0)$\\[3pt]
$Ri$							& $4.66$     & $0.40$  & $0.34$ & $0.79$\\
$Ri_g$							& $5.82$     & $0.42$  & $0.42$ & $0.99$\\
$t_c \, {\delta Ri}/{\delta t}$ & $2.47$     & $0.21$  & $0.42$ & $0.18$\\
$Ri_f$                          & $3.81$     & $0.64$  & $0.28$ & $0.65$\\
  \end{tabular}
  \caption{Transitional time for each observable and corresponding Richardson numbers. $t/t_c|_T$ (first column) is the transition time for the bulk, gradient (at the center-line), increment and flux Richardson number. Columns two to four report the values of $Ri$, $Ri_f \left(z=0\right)$ and $Ri_g$ at the corresponding transitional time of column one.}
  \label{tab:transition}
  \end{center}
\end{table}

We frame our results in the context of past research. By analyzing the results of \citet{akula2013effect,akula2017dynamics}, we evaluate the control parameter $Re_c$ to identify experiments that likely included a shear-dominated turbulent regime. A notable case is experiment $A2S2$, in which $Re_c \simeq 17000$. \citet{akula2017dynamics} evaluated the transitional location by analyzing the mixing layer growth rate and then estimated $Ri_g$ at the center-line at that location. We can thus compare their value of $Ri_g$ at the transition with line three, column four of Tab. \ref{tab:transition}: \citet{akula2017dynamics} obtained $Ri_g = 0.17$ in experiment $A2S2$, that is almost equal to our estimate ($Ri_g = 0.18$). Again, considering the mixing layer growth rate, experiment $A2S2$ transitions from shear-dominated to RT turbulence at $t/t_c|_T = 2.09$ that is close to the value $2.47$ showed in  Tab. \ref{tab:transition} (column one, line three). This further confirms that the cross-over time $t_c$ correctly predicts the transition from shear- to buoyancy-dominated turbulence for sufficiently high $Re_c$ (turbulence already developed in the shear-dominated phase). Moreover, $t_c$ is shown to correctly scale the time even when the shear-dominated phase is suppressed, provided that only the fully developed turbulent phase is considered.

\section{Concluding remarks}
In this work, we systematically analyzed the problem of the transition from shear-dominated to Boussinesq RT turbulence through theory and direct numerical simulations. Our phenomenological approach allows to predict a cross-over time $t_c \simeq \Delta U / \left( \beta g {\Delta \theta} \right)$ at which an initially shear-dominated turbulent flow transitions to RT turbulence. DNSs confirmed the validity of our theory, in particular in terms of the flux Richardson number, that is a direct measure of which factor dominates the turbulence (buoyancy or shear). Moreover, using the Buckingham-$\Pi$ theorem, we clarified in which conditions the shear-dominated turbulence is suppressed (low values of $Re_c$). This latter aspect is particularly noteworthy, since it highlights the fact that, only when the turbulence is fully developed at early times, it is reasonable to expect the flow to show a universal behavior (i.e. the transition at a unique non-dimensional time).

In Sec. \ref{sec:Results  and discussion}, we derive a scaling law for the flux Richardson number that is in good agreement with the data. This follows from the non-trivial scaling $\overline{u'w'} \sim \Delta U \left( \beta g \theta \right) t$, i.e. the Reynolds stress tensor remains anisotropic also for longer times. This scaling is expected to have implications for turbulence modeling, since in pure RT turbulence $\overline{u'w'}$ is always equal to zero. For instance, it follows from our framework that the eddy viscosity is not constant in time and should feature an explicit dependence on the growing impact of stratification also for later times when the unstable stratification dominates the turbulence.

At the end of Sec. \ref{sec:Results  and discussion} we framed our work in the context of past research. By comparing our results with the experiment by \citet{akula2017dynamics}, we observe that, at large $Re_c$, there may be a well defined value of $t/t_c|_T$ (about $2.5$), as well as of the transitional gradient Richardson number (about $0.18$). In other words, while for small values of $Re_c$ we expect a more complex dependency of $t/t_c|_T$ and $Ri_g(t/t_c|_T,z=0)$ on the control parameter $Re_c$ (and eventually on the nature of the perturbation), at finite but large $Re_c$, a unique and universal value of $t/t_c|_T$ (or $Ri_g(t/t_c|_T,z=0)$) may exist. This fact is crucial, given that the Reynolds numbers are usually large in many natural environments as well as in industrial applications. The detailed study of this aspect is left to future work.

\section*{Declaration of interests}
The authors report no conflict of interest.

\bibliographystyle{jfm}

\end{document}